\documentclass[a4paper,fleqn]{cas-sc}
\usepackage[numbers]{natbib}
\usepackage{array}
\usepackage{caption,setspace}
\usepackage{graphicx}
\usepackage{subfigure} 
\begin{document}

\title [mode = title]{On the Robustness of ``Robust reversible data hiding scheme based on two-layer embedding strategy''}                      
\author[1]{Wen Yin}
\author[1]{Longfei Ke}
\author[1]{Zhaoxia Yin}[orcid=0000-0003-0387-4806]
\cormark[1]
\cortext[mycorrespondingauthor]{Corresponding author. \quad yinzhaoxia@ahu.edu.cn (Z. Yin)}
\author[1]{Jin Tang}
\author[1]{Bin Luo}
\address[1]{Anhui Province Key Laboratory of Multimodal Cognitive Computation, School of Computer Science and Technology, Anhui University, 230601, P.R.China}

\begin{abstract} 
In the paper ``Robust reversible data hiding scheme based on two-layer embedding strategy'' published in INS recently, Kumar et al. proposed a robust reversible data hiding (RRDH) scheme based on two-layer embedding. Secret data was embedded into the most significant bit (MSB) planes to increase robustness, and a sorting strategy based on local complexity was adopted to reduce distortion. However, Kumar et al.'s reversible data hiding (RDH) scheme is not as robust against joint photographic experts group (JPEG) compression as stated and can not be called RRDH. This comment first gives a brief description of their RDH scheme, then analyses their scheme's robustness from the perspective of JPEG compression principles. JPEG compression will change pixel values, thereby destroying auxiliary information and pixel value ordering required to extract secret data correctly, making their scheme not robust. Next, the changes in both bit plane and pixel value ordering after JPEG compression are shown and analysed by different robustness-testing experiments. Finally, some suggestions are given to improve the robustness.
\end{abstract}




\maketitle
\thispagestyle{plain}
\section{Introduction}

For some sensitive scenarios such as medical and military images, lossless recovery of original images and accurate extraction of the secret data are necessary. To solve this issue, RDH is proposed to losslessly recover both the original image and the secret data \textcolor{blue}{\cite{shi2016reversible}}. The RDH technique reversibly embeds secret data into the original image to obtain a marked image. However, lossy procedures, such as JPEG compression, may cause irreversible damage to the marked image in some situations. The secret data can still be extracted correctly from the damaged image, even though the original image cannot be recovered reversibly, which is called RRDH \textcolor{blue}{\cite{shi2016reversible}}. 

De Vleeschouwer et al.  \textcolor{blue}{\cite{de2003circular}} introduced the first RRDH scheme, which proposed the histogram rotation technique.
Their scheme suffered from salt-and-pepper noise because of modulo-256 addition, resulting in poor image quality after embedding a watermarking. To avoid that in \textcolor{blue}{\cite{de2003circular}}, Ni et al. \textcolor{blue}{\cite{ni2008robust}} proposed an RRDH scheme using a robust parameter (difference value of all pixels in the block) to embed data and adopting error correction coding to achieve reversibility. Zeng et al. \textcolor{blue}{\cite{zeng2010lossless}} increased the robustness of the scheme \textcolor{blue}{\cite{ni2008robust}} by using the arithmetic difference of blocks as robust features and embedding secret data using two thresholds and a new histogram modification scheme. In \textcolor{blue}{\cite{gao2011lossless}}, Gao et al. improved Zeng et al.'s scheme by designing a more stable framework to increase the embedding capacity and robustness. The RRDH scheme has also been developed in the wavelet domains. Based on the characteristics of laplacian distribution of block mean values in wavelet domain, Zou et al. \textcolor{blue}{\cite{zou2006semi}} designed an RRDH scheme. Later, An et al. \textcolor{blue}{\cite{an2012content}} \textcolor{blue}{\cite{an2012robust}} improved the robustness by using histogram shifting and clustering, as well as efficiently dealing with overflow and underflow to achieve reversibility. Coltuc et al. \textcolor{blue}{\cite{coltuc2007towards}} \textcolor{blue}{\cite{coltuc2007distortion}} proposed a two-stage framework for RRDH. The first stage was the robust embedding stage for extracting the secret data accurately, and the second stage was used to restore the original image reversibly. However, because both the robust embedding step and the reversible embedding stage were operated on the same embedding domain, the watermarking extraction may fail. To address this problem,
Wang et al. \textcolor{blue}{\cite{wang2019independent}} offered an independent embedding domain scheme to preserve robustness. Coltuc et al.'s scheme was improved by splitting the original image into two distinct domains, one for robust watermarking and the other for reversible embedding. Xiong et al. \textcolor{blue}{\cite{xiong2021robust}}  proposed a multi-security protection RRDH scheme that uses patchwork robust watermarking to provide the scheme's robustness and prediction error expansion to ensure the scheme's reversibility exploiting the two-stage architecture.

Recently, Wang et al. \textcolor{blue}{\cite{wang2017reversible}} proposed a lossless RRDH scheme based on significant-bit-difference expansion, in which secret data was embedded into higher significant bit planes of the image to improve the embedding capacity. Inspired by the work of \textcolor{blue}{\cite{wang2017reversible}}, Kumar et al. \textcolor{blue}{\cite{kumar2020robust}} first decomposed the original image into two planes, MSB (Here MSB is equivalent to HSB in \textcolor{blue}{\cite{kumar2020robust}}, but since the more general term is MSB, we use MSB instead.) planes and least significant bit (LSB) planes. Then, the MSB planes of the original image were used to embed secret data by making a prediction error expansion with two-layer embedding. In this way, Kumar et al. achieved the robustness of their scheme against minor modification attacks like JPEG Compression. However, in this comment, we demonstrate that the scheme \textcolor{blue}{\cite{kumar2020robust}} is not robust. The contributions of this paper are summarized as follows: \\
\begin{enumerate}[(1)]
\item This paper tests the percentage of bits changed in each bit plane after JPEG compression and finds that JPEG compression will also cause the contents of MSB planes to be damaged.\\
\item This paper investigates the scheme \textcolor{blue}{\cite{kumar2020robust}} and analyses the changes in both each bit plane and pixel value ordering caused by JPEG compression to show that secret data can not be extracted correctly, thus proving that their scheme is not robust.\\
\item Some suggestions are given to improve the robustness of the scheme \textcolor{blue}{\cite{kumar2020robust}} from different perspectives.
\end{enumerate}

The rest of the paper is organized as follows. In \textbf{Section 2}, the scheme \textcolor{blue}{\cite{kumar2020robust}} is briefly reviewed. In \textbf{Section 3}, a theoretical analysis is conducted from the perspective of JPEG compression principles. Experimental results and analysis are shown in \textbf{Section 4}. In \textbf{Section 5}, some suggestions to improve the robustness of the scheme \textcolor{blue}{\cite{kumar2020robust}} are given. Finally, the paper is concluded in \textbf{Section 6}.

\section{Brief description of Kumar et al.'s RDH scheme \textcolor{blue}{\cite{kumar2020robust}}}

For a self-contained discussion, we briefly review the scheme \textcolor{blue}{\cite{kumar2020robust}} and recommend \textcolor{blue}{\cite{kumar2020robust}} to readers for more detailed information about it.

To facilitate the introduction of the scheme \textcolor{blue}{\cite{kumar2020robust}}, secret data is denoted as $\boldsymbol{S}$, and an original image of size $h \times w$ is denoted as $\boldsymbol{I}$. The pixel values of $\boldsymbol{I}$ vary in the range of 0-255. The eight main steps of the embedding scheme are summarized as follows:

\begin{enumerate}[\ \ \ \ Step 1:]

\item Divide $\boldsymbol{I}$ into two parts $\boldsymbol{I}_{MSB}$ and $\boldsymbol{I}_{LSB}$ calculated as
\begin{normalsize}
\begin{equation}
p_{i,j} = x_{i,j} + l_{i,j}, \quad i=0,...,h-1,\quad j=0,...,w-1, \label{1}
\end{equation}
\end{normalsize}
where
\begin{normalsize}
\begin{equation}
x_{i,j} = \sum_{k=n}^{7}b_{k}\times 2^{k}, \label{2}
\end{equation}
\end{normalsize}
\begin{normalsize}
\begin{equation}
l_{i,j} = \sum_{k=0}^{n-1}b_{k}\times2^k. \label{3}
\end{equation}
\end{normalsize}\\
$p_{i,j}$ denotes the pixel value at the coordinate $(i,j)$ of $\boldsymbol{I}$, \textit{$x_{i,j}$} represents the pixel value in MSB planes of the pixel, \textit{$l_{i,j}$} denotes the pixel value in LSB planes of the pixel, and $b_{k}$ is the bit value in the \textit{k}-th location. $n$ represents the number of planes of LSB planes, $n$ varies from 1 to 8.

\item Preprocess $\boldsymbol{I}_{MSB}$, construct a location map and compress it to achieve a compressed location map. 

\item Save the least significant bit of pixels of $\boldsymbol{I}_{MSB}$ and attach the saved bits and compressed location map to the end of $\boldsymbol{S}$.

\begin{figure}[h]
\centering
\subfigure[]{
\includegraphics[width=5cm]{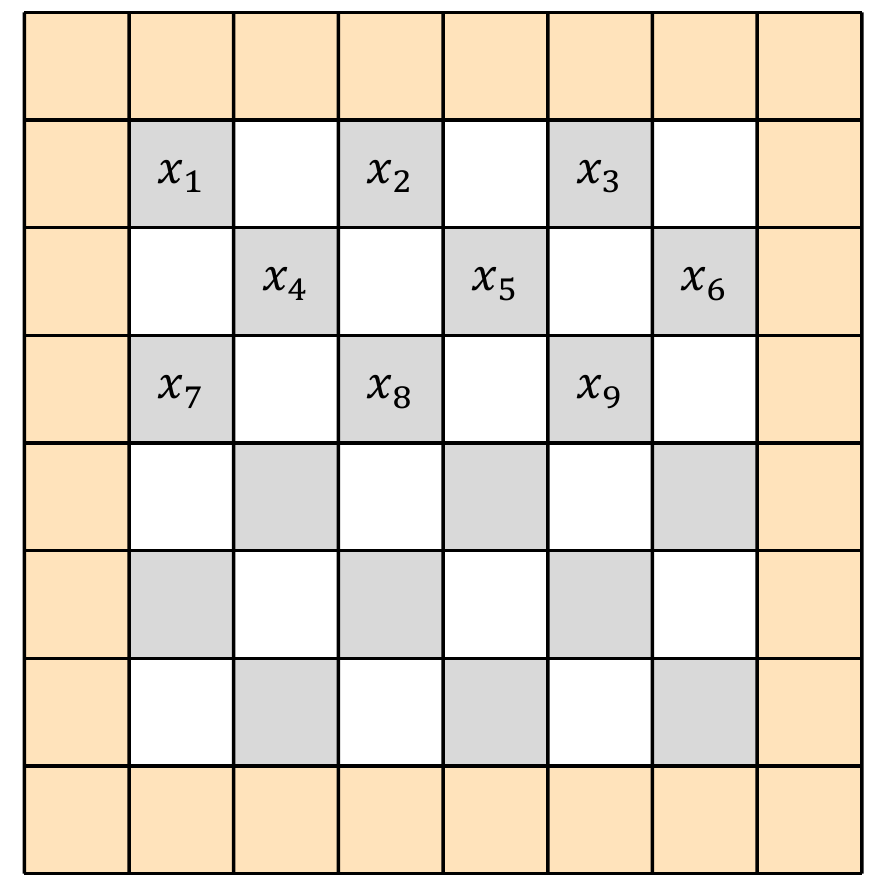}
}
\quad
\subfigure[]{
\includegraphics[width=5cm]{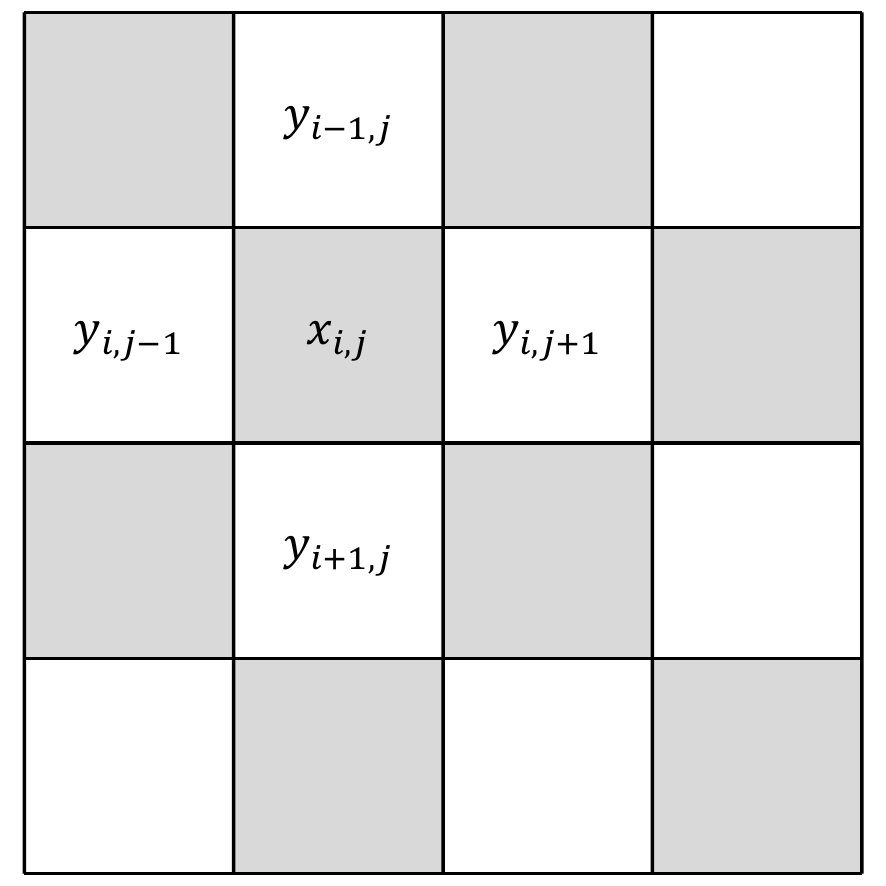}
}
\quad
\caption{\centering \textrm{(a) Chessboard pattern; (b) Prediction pattern.}}
\vspace{-0.2cm}
\end{figure}

\item Make use of the pattern in \textcolor{blue}{Fig. 1(a)} to divide the image as in \textcolor{blue}{\cite{kumar2020robust}}. Sort the grey pixels by local complexity and use two-layer embedding  in \textcolor{blue}{\cite{wang2017reversible}} to embed the first part of \textbf{\textit{S}}. The local complexity of $x_{i,j}$ is calculated according to variance ($\mu_{i,j}$) of surrounding pixels ($y_{i,j-1}$,$y_{i-1,j}$,$y_{i,j+1}$,$y_{i+1,j}$)(see \textcolor{blue}{Fig. 1(b)}) by
\begin{normalsize} 
\begin{equation}
\mu_{i,j}=\frac{1}{4}\sum_{t=1}^{4}\left ( M_{i,j}-v_{t} \right )^2, \label{4}
\end{equation}
\end{normalsize}\\
where $v_{t} (1\leq t\leq 4)$ is the ascending sequence of the surrounding pixels,  $M_{i,j}$ is the mean value of $v_{t} (1\leq t\leq 4)$.

\item Sort the white pixels based on their local complexity, and the remaining part of \textbf{\textit{S}} is embedded by using prediction error expansion and two-layer embedding. The first prediction error ($e_{1}$) is calculated by $e_{1}=x_{i,j}-\hat{p_{1}}$ where $\hat{p_{1}}=\left \lfloor \frac{v_{1}+v_{2}+v_{3}}{3} \right \rfloor$ if $N=3$ and $\left \lfloor \cdot  \right \rfloor$ represents floor operators. $N$ is the predictor number. The secret data bit $s_{1}\in \left \{ 0,1 \right \}$ is embedded using
\begin{normalsize} 
\begin{equation}
x_{i,j}^{'}=\left\{\begin{matrix}
x_{i,j}+s_{1},& \quad e_{1}=1, & \\[0.2cm] 
x_{i,j}+1, & \quad e_{1}>1, & 
\\[0.2cm] 
x_{i,j},& \quad e_{1}<1, & 
\end{matrix}\right.\label{5}
\end{equation}
\end{normalsize}\\
where $x_{i,j}^{'}$ is the obtained pixel value after the first layer embedding. The second prediction error ($e_{2}$) is calculated by $e_{2}=x_{i,j}^{'}-\hat{p_{2}}$ where $\hat{p_{2}}=\left \lfloor \frac{v_{2}+v_{3}+v_{4}}{3} \right \rfloor$ if $N=3$. The secret data bit $s_{2}\in \left \{ 0,1 \right \}$  is embedded using 

\begin{normalsize} 
\begin{equation}
x_{i,j}^{''}=\left\{\begin{matrix}
x_{i,j}^{'}-s_{2}, & \quad e_{2}=-1, & 
\\[0.2cm] 
x_{i,j}^{'}-1, & \quad e_{2}<-1, & 
\\[0.2cm] 
x_{i,j}^{'}, & \quad e_{2}>-1, &
\end{matrix}\right.\label{6}
\end{equation}
\end{normalsize}\\
where $x_{i,j}^{''}$ is the obtained pixel value after the second layer embedding.

\item Combine the resultant MSB planes and LSB planes to obtain the marked image.

\item Repeat Step 4 to Step 6 for all the three predictors, select the most suitable predictor $(N)$ and the marked image according to application requirements.

\item Replace the ($n$+1)-th least significant bit of border pixels of the marked image to save auxiliary information ($N$ and the coordinate of the last pixel to embed the secret data ($C_{end}$)) using the least significant bit substitution scheme. 

\end{enumerate}

\section{Theoretical analysis from JPEG compression principles}

Since Kumar et al. \textcolor{blue}{\cite{kumar2020robust}} only mentioned the robustness of their scheme against JPEG compression, we will focus on analysing robustness against JPEG compression. This section first briefly reviews the JPEG compression process and then analyses how this process changes the pixel value and destroys the robustness of the scheme \textcolor{blue}{\cite{kumar2020robust}}.\\

\noindent 3.1. Overview of JPEG compression\\

 JPEG compression is the most widely used lossy image compression method on the Internet. It makes use of the characteristics of the human visual system and uses the combination of quantization and lossless compression coding to remove the redundant information of the original image itself. \textcolor{blue}{Fig. 2} shows the JPEG compression process, which is executed by three parts, namely Discrete Cosine Transform (DCT), Quantizer, Entropy encoder \textcolor{blue}{\cite{wallace1992jpeg}}.

\begin{figure}[h]
\centering
	\includegraphics[scale=.4]{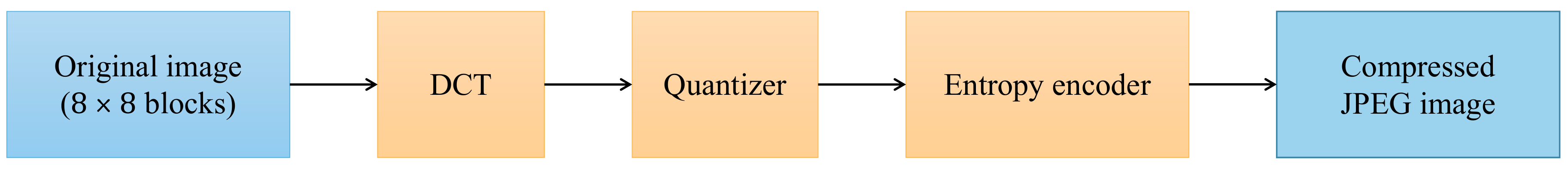}
\centering
\caption{\centering \textrm{JPEG compression encoding process.}}

\end{figure}

Applying two-dimensional DCT transformation to the non-overlapping $8\times 8$ blocks, the original DCT Coefficients are obtained by  
\begin{normalsize} 
\begin{equation}
\qquad o\left ( u,v \right )=\frac{1}{4}\alpha \left ( u \right )\alpha \left ( v \right )\sum_{i=0}^{7}\sum_{j=0}^{7}p_{i,j}{\rm cos}\frac{\left ( 2i+1 \right )u{\rm \pi}}{16}{\rm cos}\frac{\left ( 2j+1 \right )v{\rm \pi} }{16},
u=0,...,7, v=0,...,7,
\end{equation}
\end{normalsize}\\
where 
\begin{normalsize} 
\begin{equation}
\qquad \alpha \left ( u \right )= \left\{\begin{matrix}
\frac{1}{\sqrt{2}}, \quad u=0,
\\[0.2cm] 
1,  \quad else,
\end{matrix}\right. 
\end{equation}
\end{normalsize}\\
and $p_{i,j}$ represents the pixel value of a block at the position $(i,j)$ and $o(u,v)$ is the original DCT coefficient at the position $(u,v)$ of the block. Then the original image in the spatial domain is converted into an image in the frequency domain.

Next, the original DCT coefficients are quantized by a quantizer, which is the main reason for image quality deterioration. In the quantizer, the original DCT coefficients are processed by
\begin{normalsize} 
\begin{equation}
\qquad r\left ( u,v \right )={\rm round} \left ( \frac{o\left ( u,v \right )}{q\left ( u,v \right )} \right ),
\end{equation}
\end{normalsize}\\
where $r(u,v)$ is the quantized DCT coefficient, $q(u,v)$ is the predetermined quantization step of the position $(u,v)$ in the quantization tables for different quality factors (QF), round($t$) means round to nearest integer to $t$. Through this step, the floating values of the original DCT coefficients are rounded to an integer, resulting in the loss of information, which is irreversible. It can be seen from \textcolor{blue}{Eq. (9)} that the larger the quantization step $q(u,v)$ is, the larger error is introduced by the rounding process. Depending on the compression ratio of JPEG images, different quantization tables can be selected. Generally speaking, the larger the compression rate (that is, the smaller the QF), the larger the quantization step in the quantization table, as shown in \textcolor{blue}{Fig. 3}. As can be seen, the quantization step of the quantization table with QF = 70 is larger. Therefore, compared with QF = 80, the pixel value after JPEG compression with QF = 70 is tempered with more seriously. Next, this variation  is visualized to analyse the robustness of the scheme \textcolor{blue}{\cite{kumar2020robust}}.\\

\begin{figure}[h]
\centering
\subfigure[]{
\includegraphics[width=5cm]{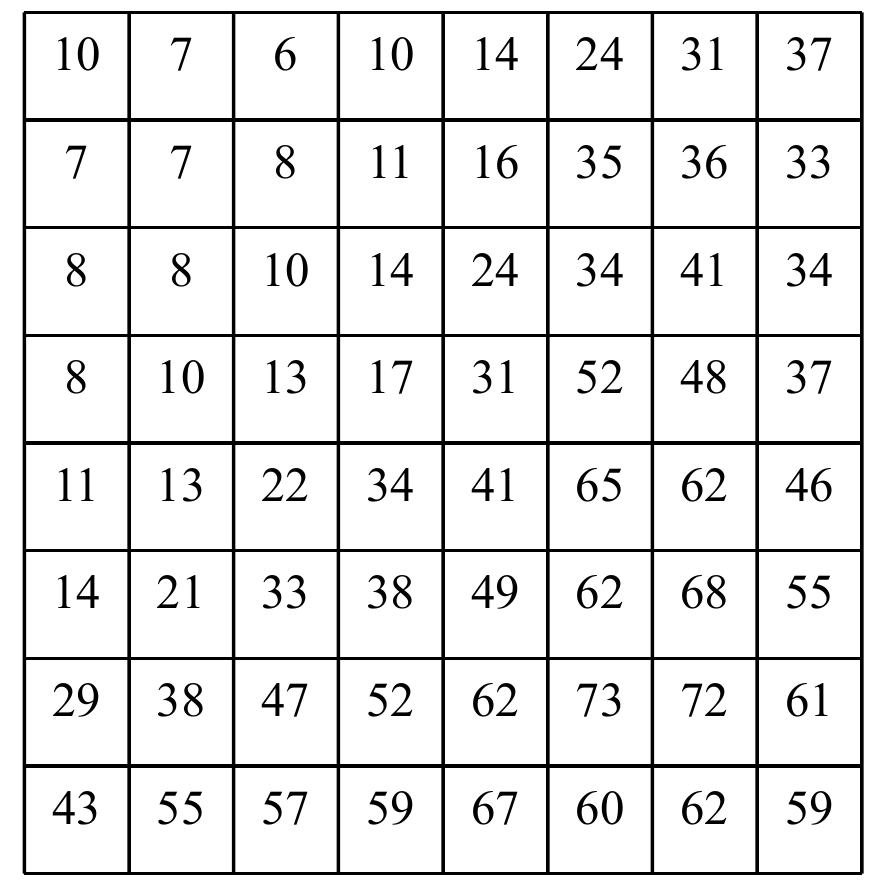}
}
\quad
\subfigure[]{
\includegraphics[width=5cm]{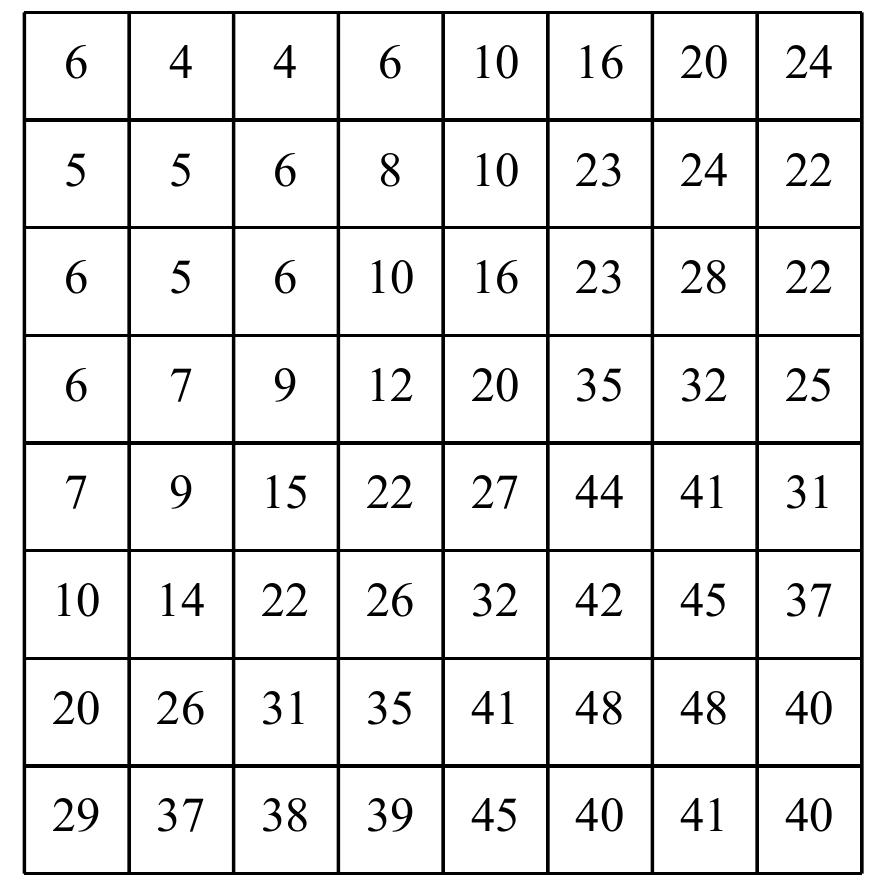}
}
\quad
\caption{\centering \textrm{Quantization tables: (a)  QF = 70; (b) QF = 80.}}

\end{figure}

\noindent 3.2. Pixel value variation caused by JPEG compression\\

We randomly select an $8\times8$ block of an image in Bossbase 1.01 \textcolor{blue}{\cite{bas2011break}} to calculate the change of pixel values before and after JPEG compression, which is visualized in \textcolor{blue}{Fig. 4}. \textcolor{blue}{Fig. 4(b)} and \textcolor{blue}{(c)} illustrate the absolute values of the modifications on the compressed pixels compared to the original when QF = 70 and QF = 80 respectively.    As we can see, QF = 70 causes a larger variation in pixel values than QF = 80, and the like, with  QF decreases, this variation will be more and more obvious. The reason why pixel values vary so much is that, from \textcolor{blue}{Eq. (7)}, one DCT coefficient of a block is calculated by all pixel values of the block in the spatial domain. Therefore, the pixel values will be severely tampered with after rounding  all DCT coefficients in the block, which is irreversible. As shown in \textcolor{blue}{Fig. 4(b)}  and \textcolor{blue}{(c)}, the pixel values required to extract secret data accurately will be greatly modified after JPEG compression, which will lead to errors when the receiver extracts secret data. Next, we will show robust testing experiments for the scheme \textcolor{blue}{\cite{kumar2020robust}} to illustrate how variation in pixel values affects the robustness of their scheme.

\begin{figure}[h]
\centering
\subfigure[]{
\includegraphics[width=4.6cm]{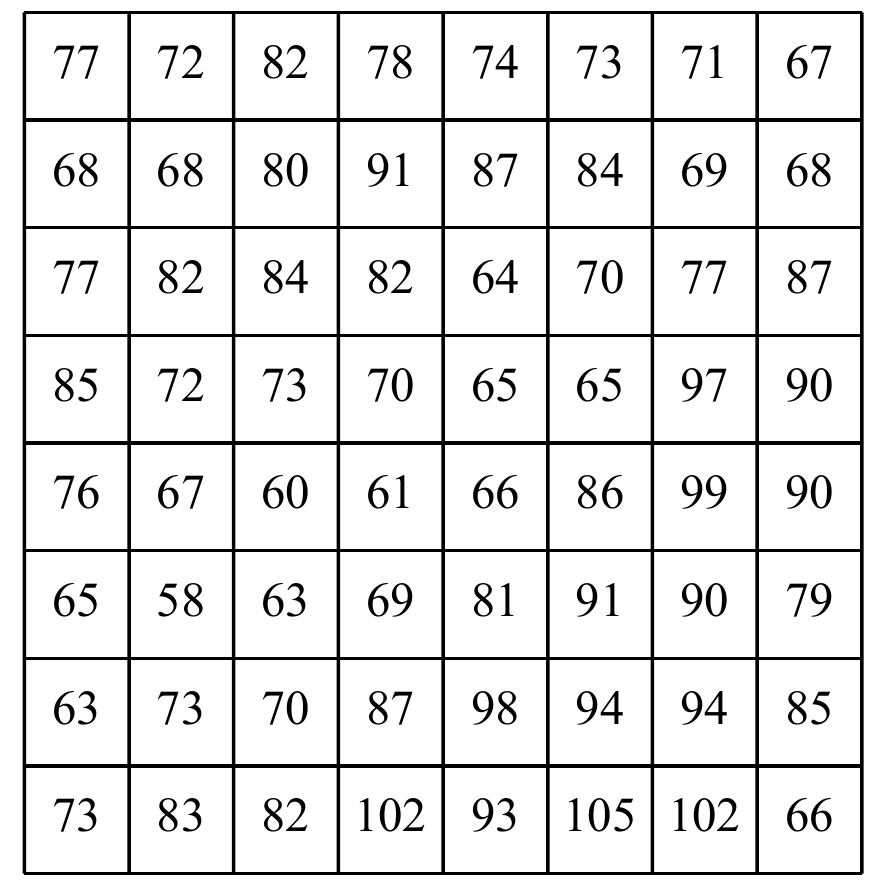}
}
\subfigure[]{
\includegraphics[width=5cm]{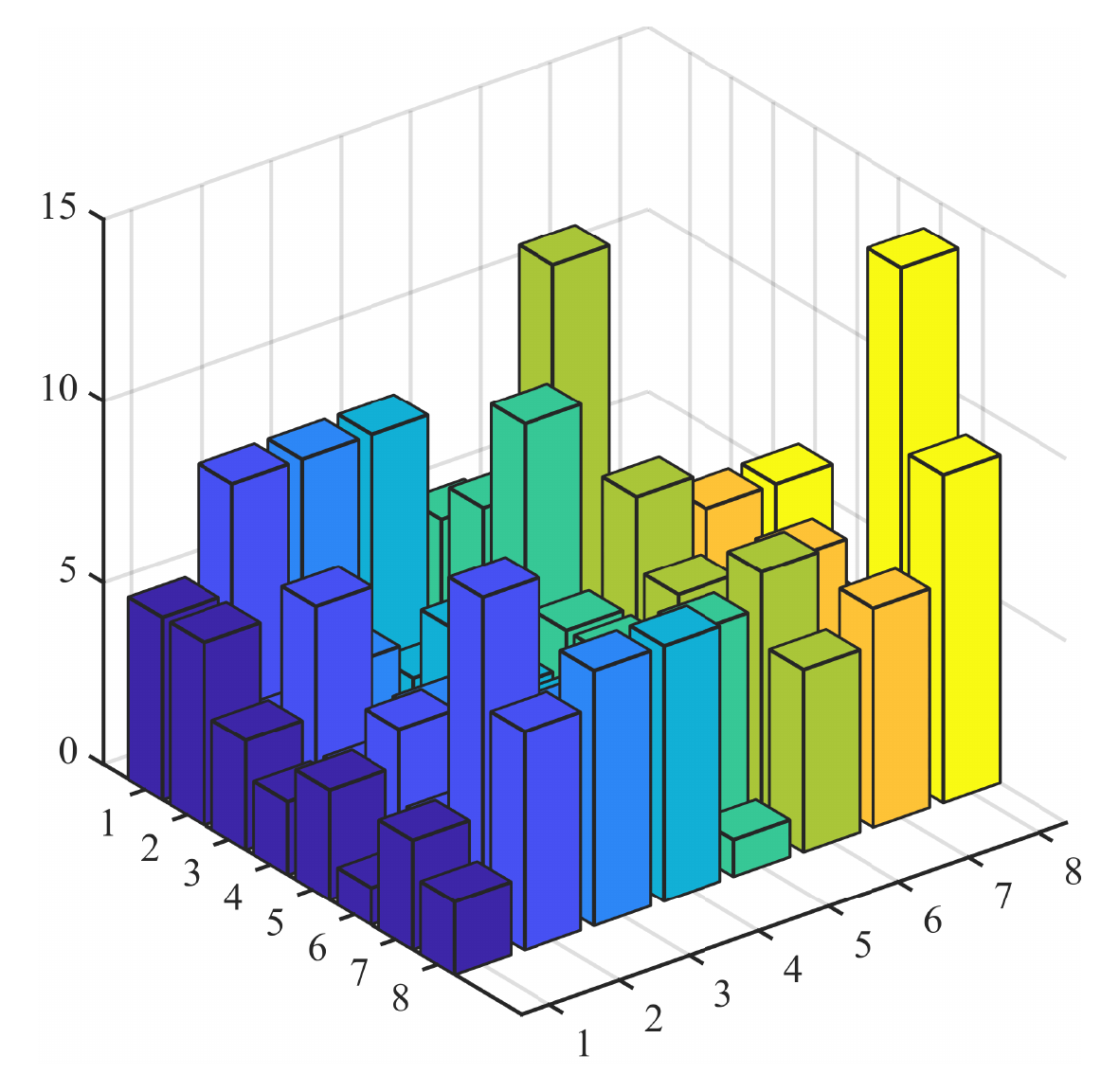}
}
\subfigure[]{
\includegraphics[width=5cm]{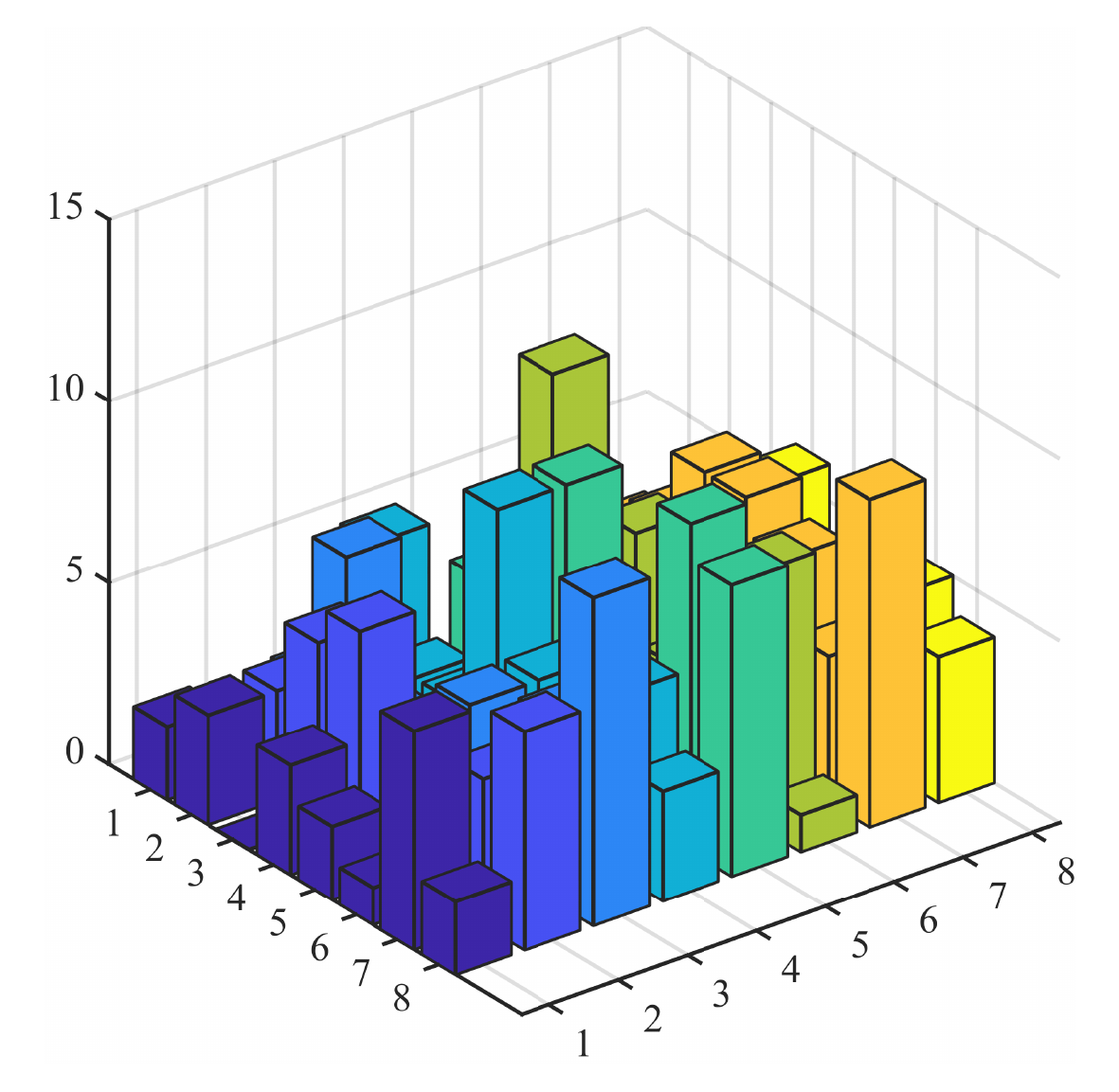}
}
\quad
\caption{\centering \textrm{Effect of different quantization tables on pixel values: (a) Pixel block; (b) QF = 70; (c) QF = 80.}}

\end{figure}

\section{Robust testing against JPEG compression}

As described in Step 4 of Section 2, the scheme \textcolor{blue}{\cite{kumar2020robust}} used the partition pattern in \textcolor{blue}{Fig. 1(a)} to generate independent cells, then rearranged the grey pixels with ascending order according to local complexity. In this way, secret data can be embedded into pixels with lower local complexity to achieve less distortion. To extract the data correctly, the receiver also sorts the pixels on the basis of local complexity and then processes them in order.\\

\noindent 4.1. Changes in MSB planes caused by JPEG compression\\

Kumar et al. \textcolor{blue}{\cite{kumar2020robust}} showed that embedding secret data in MSB planes can improve the robustness of the scheme as slight attacks like JPEG compression make modifications in the lower bit planes, so the content of MSB planes is intact. In fact, JPEG compression results in image pixels changing not only in LSB planes but also in MSB planes. The experiments were carried out on eight standard grey-scale images used by Kumar et al. as shown in \textcolor{blue}{Fig. 5}. Each of size $512 \times 512$ pixels including Lena, Baboon, Jetplane, Peppers, Barbara,  Lake, Elaine and Boat.

\newcommand{\mysize}{2.4cm}
\begin{figure}[h]
\centering
\subfigure[]{
\includegraphics[width=\mysize]{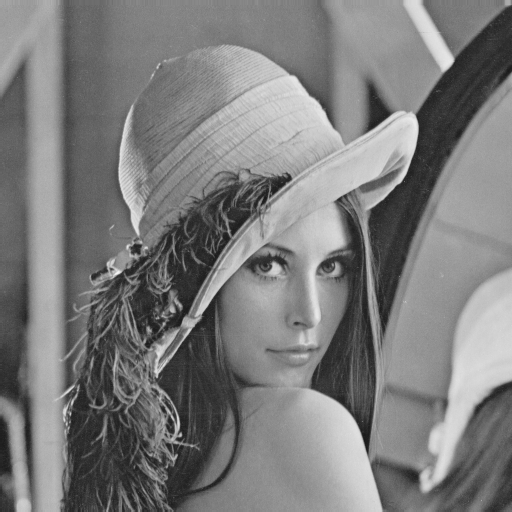}
}
\quad
\subfigure[]{
\includegraphics[width=\mysize]{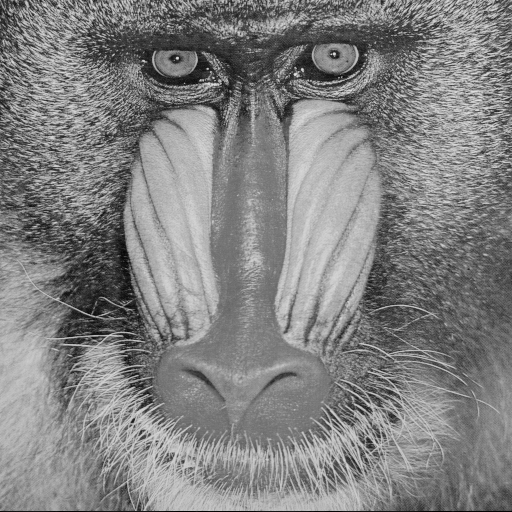}
}
\quad
\subfigure[]{
\includegraphics[width=\mysize]{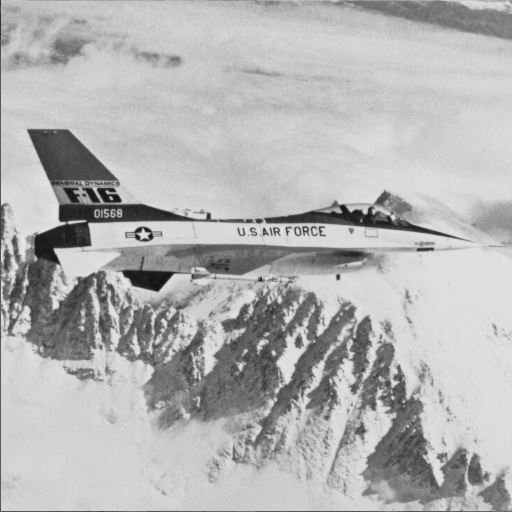}
}
\quad
\subfigure[]{
\includegraphics[width=\mysize]{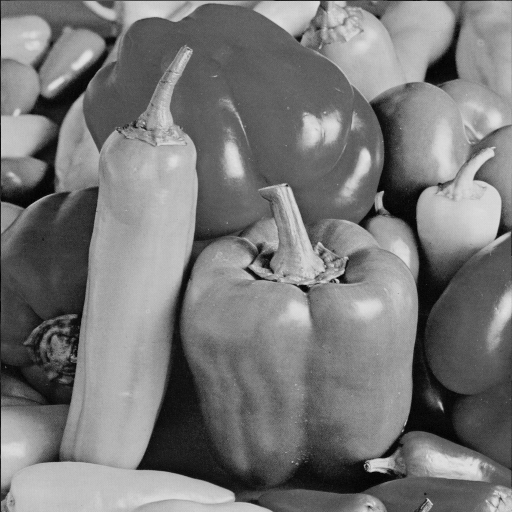}
}
\quad\\

\subfigure[]{
\includegraphics[width=\mysize]{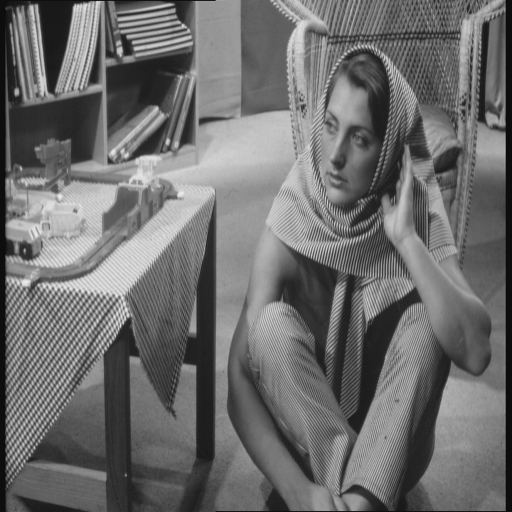}
}
\quad
\subfigure[]{
\includegraphics[width=\mysize]{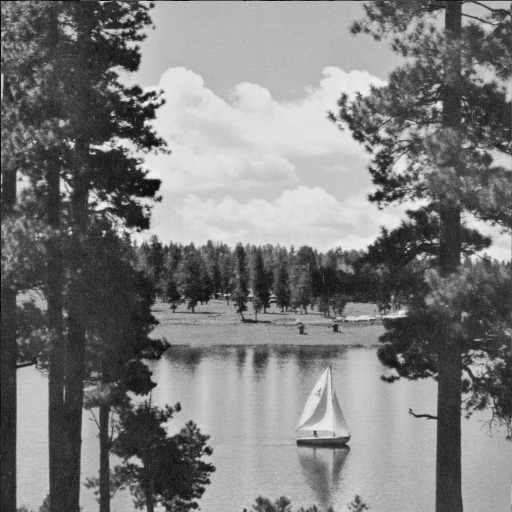}
}
\quad
\subfigure[]{
\includegraphics[width=\mysize]{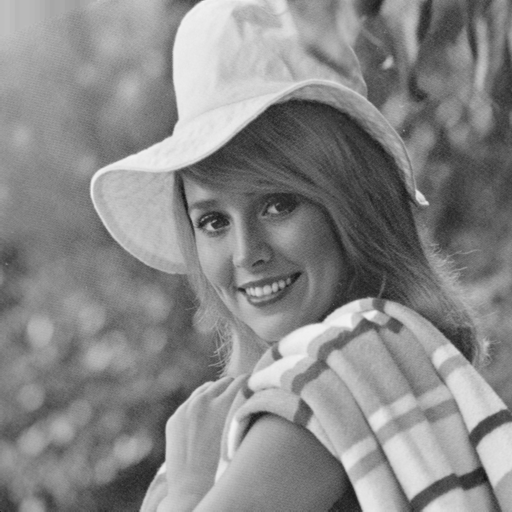}
}
\quad
\subfigure[]{
\includegraphics[width=\mysize]{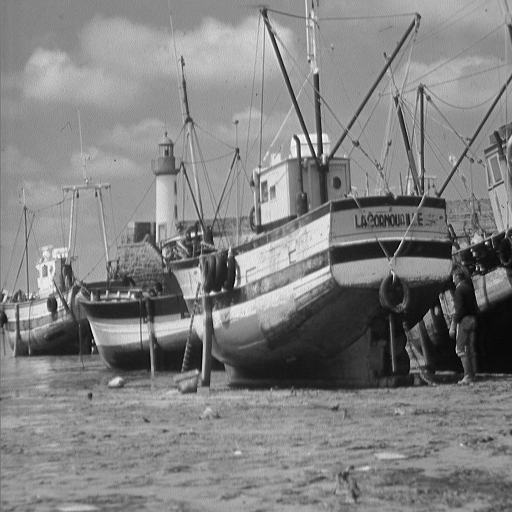}
}
\caption{\textrm{Original images: (a) Lena; (b) Baboon; (c) Jetplane; (d) Peppers; (e) Barbara; (f) Lake; (g) Elaine; (h) Boat.}}
\vspace{-0.1cm}
\end{figure}

We perform JPEG compression  with different QF on the original images. The dissimilarity, between the original image ($\boldsymbol{I}$) and the processed image ($\boldsymbol{I}'$) after JPEG compression, has been observed by the number of bit change rate (NBCR). The NBCR indicates the percentage of bits in each bit plane changed after JPEG compression. NBCR is formulated as 
\begin{small}
\begin{equation}
\qquad {\rm NBCR}=\frac{1}{h\times w}\left [ \sum_{i=1}^{h}\sum_{j=1}^{w} \left ( \boldsymbol{I}_k\left ( i,j \right ) \oplus \boldsymbol{I}'_k\left ( i,j \right ) \right ) \right] \times 100\%,
\end{equation}
\end{small}\\
where $\oplus$ represents exclusive-or (XOR) operation, $\boldsymbol{I}_k\left (i,j\right)$ and $\boldsymbol{I}'_k\left ( i,j \right )$ denote the position $(i,j)$ of the $k$-th bit plane of the image $\boldsymbol{I}$ and $\boldsymbol{I}'$ respectively.

The testing results are shown in \textcolor{blue}{Fig. 6}, where the horizontal axis represents different QF, the vertical axis represents the average NBCR of eight original images, the 1-st bit plane represents the least significant bit plane, and the 8-th bit plane represents the most significant bit plane. We can see that the NBCR of the 1-st bit plane reaches about 50\%  when the quality factor is below 95, which indicates that the least significant bit plane can be easily changed. Among eight bit planes, the 8-th bit plane has the lowest NBCR, which is the most stable. But even at the lowest NBCR, the average NBCR of eight images is 0.0574\%. For each bit plane, as the QF increases, NBCR decreases, which implies that the larger the QF, the fewer bits are modified. Therefore, next, we will select QF = 100 and QF = 95 for further analysis. 

\begin{figure}[h]
\centering
	\includegraphics[scale=.45]{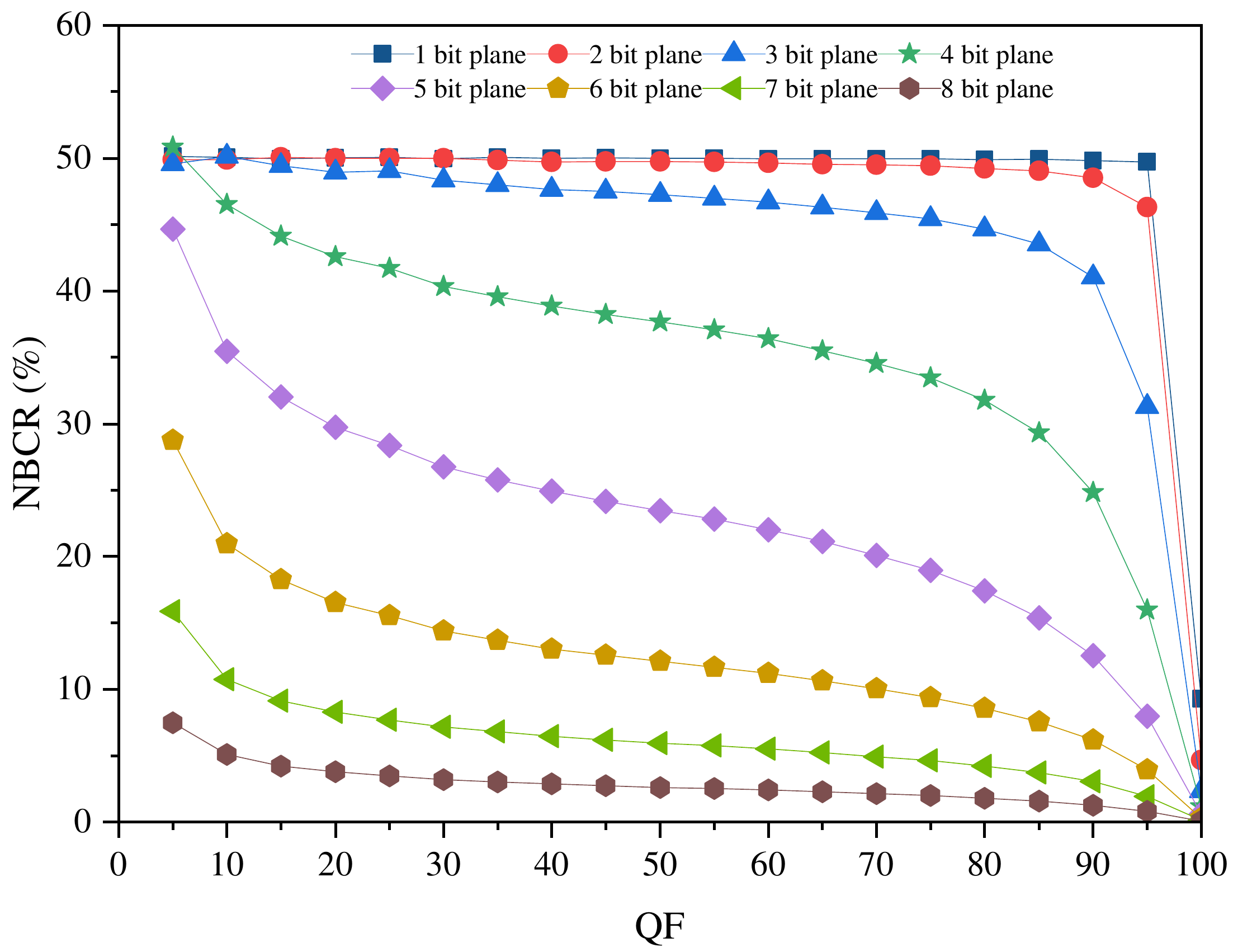}
\centering
\caption{\centering \textrm{Average NBCR of 8 original images on different bit planes.}}
\vspace{-0.5cm}
\end{figure}

\begin{table}[width=.9\linewidth,cols=4,pos=h]
\caption{\textrm{NBCR (\%) in each bit plane after JPEG compression (QF = \textrm{100}).}}\label{tbl1}
\begin{tabular*}{\tblwidth}{@{} LCCCCCCCC@{} }
\toprule
\textrm{Bit Plane} & \textrm{8} & \textrm{7} & \textrm{6} & \textrm{5} & \textrm{4} & \textrm{3} & \textrm{2} & \textrm{1} \\
\midrule
\textrm{Lena} & \textrm{0.0702} & \textrm{0.1335} & \textrm{0.2872} & \textrm{0.6012} & \textrm{1.1467} & \textrm{2.3182} & \textrm{4.6494} & \textrm{9.2121} \\
\textrm{Baboon} & \textrm{0.0870} & \textrm{0.1606} & \textrm{0.2853} & \textrm{0.5722} & \textrm{1.1600} & \textrm{2.3518} & \textrm{4.6864} & \textrm{9.2445} \\
\textrm{Jetplane} & \textrm{0.0244} & \textrm{0.1183} & \textrm{0.2285} & \textrm{0.5699} & \textrm{1.1272} & \textrm{2.3182} & \textrm{4.6581} & \textrm{9.2857} \\
\textrm{Peppers} & \textrm{0.0401} & \textrm{0.1366} &	\textrm{0.3155} & \textrm{0.6081} & \textrm{1.1559}	& \textrm{2.3052} &	\textrm{4.6494} & \textrm{9.2991} \\
\textrm{Barbara} & \textrm{0.0496} & \textrm{0.1251} & \textrm{0.3006} & \textrm{0.5993} & \textrm{1.1227} & \textrm{2.2797} & \textrm{4.6474} & \textrm{9.3307} \\
\textrm{Lake} & \textrm{0.0210} & \textrm{0.1514} &	\textrm{0.3185} & \textrm{0.5810} &	\textrm{1.1532} & \textrm{2.2892} &	\textrm{4.6082} & \textrm{9.2186} \\
\textrm{Elaine} & \textrm{0.0839} & \textrm{0.1637} & \textrm{0.2968} &	\textrm{0.5867} & \textrm{1.1631}	& \textrm{2.3140} &	\textrm{4.6677 }& \textrm{9.2991} \\
\textrm{Boat }& \textrm{0.0832 }& \textrm{0.1217} &	\textrm{0.2831 }& \textrm{0.5165 }&	\textrm{1.1917} & \textrm{2.5196 }&	\textrm{4.7157} & \textrm{9.2869} \\
\midrule
\textrm{Average} & \textrm{0.0574} & \textrm{0.1389} &	\textrm{0.2894} &	\textrm{0.5794 }&	\textrm{1.1526 }&	\textrm{2.3370} &	\textrm{4.6603} &	\textrm{9.2721}  \\
\bottomrule
\end{tabular*}
\vspace{-0.2cm}
\end{table}

\begin{table}[width=.9\linewidth,cols=4,pos=h]
\caption{\textrm{NBCR (\%) in each bit plane after JPEG compression (QF = 95).}}\label{tb2}
\begin{tabular*}{\tblwidth}{@{} LCCCCCCCC@{} }
\toprule
\textrm{Bit Plane} & \textrm{8} & \textrm{7}& \textrm{6}& \textrm{5}& \textrm{4}& \textrm{3} & \textrm{2} & \textrm{1}\\
\midrule
\textrm{Lena}& \textrm{1.1330}& \textrm{1.9867} &	\textrm{3.8849} &\textrm{8.2031} &	\textrm{15.6834} & \textrm{30.9162} & \textrm{46.8094} & \textrm{49.9676} \\
\textrm{Baboon} & \textrm{1.3954}&	\textrm{2.7122} & \textrm{4.7249} &	\textrm{9.3334 }& \textrm{18.7298} & \textrm{36.0718} & \textrm{48.8811} &	\textrm{49.8756} \\
\textrm{Jetplane} & \textrm{0.3490} & \textrm{1.5476} & \textrm{2.5654}& \textrm{6.1642} & \textrm{12.4996}&	\textrm{25.2193}&	\textrm{41.7076}& \textrm{48.8037} \\
\textrm{Peppers} & \textrm{0.5638} & \textrm{1.9424} &	\textrm{4.7485} & \textrm{9.1179} &	\textrm{17.7017} & \textrm{34.4444}& \textrm{48.5249} & \textrm{49.9241} \\
\textrm{Barbara} & \textrm{0.6072} &	\textrm{1.3199} &	\textrm{3.2070} &	\textrm{6.3507} &	\textrm{12.2925} &	\textrm{24.3813} &	\textrm{41.9697} &\textrm{49.5552}\\
\textrm{Lake}& \textrm{0.2911} &\textrm{ 2.2320 }&	\textrm{4.3751} & \textrm{8.7959 }&\textrm{18.0351} & \textrm{34.5795 }&\textrm{48.0885} & \textrm{50.0702} \\
\textrm{Elaine}& \textrm{1.1753} &	\textrm{2.1294} &	\textrm{3.8651}	& \textrm{7.5867}&	\textrm{15.3828} & 	\textrm{30.1125} & 	\textrm{45.9869}&\textrm{49.5667}\\
\textrm{Boat} & \textrm{1.0132}& \textrm{1.5816} &	\textrm{4.0531} & \textrm{8.1760} &	\textrm{17.3714 }&	\textrm{34.3925} &	\textrm{48.3948} &	\textrm{50.0286} \\
\midrule
\textrm{Average}& \textrm{0.8160} &	\textrm{1.9315} &	\textrm{3.9280} &	\textrm{7.9660} &	\textrm{15.9620} &	\textrm{31.2647} &	\textrm{46.2954} &	\textrm{49.7240 } \\
\bottomrule
\end{tabular*}
\vspace{-0.4cm}
\end{table}

As described in Step 8 in Section 2, in the scheme \textcolor{blue}{\cite{kumar2020robust}}, auxiliary information necessary to accurately extract secret data is embedded into the $(n+1)$-th LSB. Therefore, if we set $n=3$ as an example, the auxiliary information will be stored in the 4-th bit plane, so we will focus on the 4-th bit plane. From \textcolor{blue}{Table 1}, the 4-th bit plane where the auxiliary information is embedded has an average NBCR of 1.1526\% after JPEG compression under QF = 100. Therefore, auxiliary information can be damaged easily. The average NBCR of MSB and LSB are 0.4437\% and 5.4231\%, respectively, so neither MSB nor LSB planes are robust. From \textcolor{blue}{Table 2}, when the compression QF is 95, the average NBCR of the 4-th bit plane has reached 15.9620\%. The average NBCR of MSB and LSB reached 6.1207\% and 42.4280\%, respectively. Obviously, as the compression QF decreases, the possibility of MSB being destroyed increases. So the auxiliary information is more likely to be damaged. The length of the extracted data can be significantly wrong due to the damaged $C_{end}$. With the damaged $N$, the prediction method will be wrong.  Minor modifications will also have a huge impact on the entire scheme. As a result, the receiver will be unable to extract secret data accurately.\\

\noindent 4.2. Changes in pixel value ordering 
caused by JPEG compression\\

In addition, in Step 4 in Section 2, before embedding the secret data, the scanned pixels are arranged according to the order of their local complexity (i.e., ascending order). However, according to the calculation formula of local complexity as shown in \textcolor{blue}{Eq. (4)}, changes in pixel value will cause errors of the local complexity value, which affect the arrangement order of the pixel. Thus, JPEG compression will cause that the secret data cannot be extracted correctly. The experimental results of the pixel sequence sorted by local complexity before and after compression are shown in \textcolor{blue}{Figs. 7-9} and \textcolor{blue}{Table 3}. To demonstrate the change of pixel ordering clearly, we randomly select an $ 8\times8 $ pixel block from Lena, as shown in \textcolor{blue}{Fig. 7(a)}. The matrix of the selected area is then divided into two parts by \textcolor{blue}{Eqs. (2)} and \textcolor{blue}{(3)} namely MSB and LSB matrix under $n=3$. The MSB matrix is shown in \textcolor{blue}{Fig. 7(b)}.

\begin{figure}[htbp]
\centering
\subfigure[]{
\includegraphics[scale=0.7]{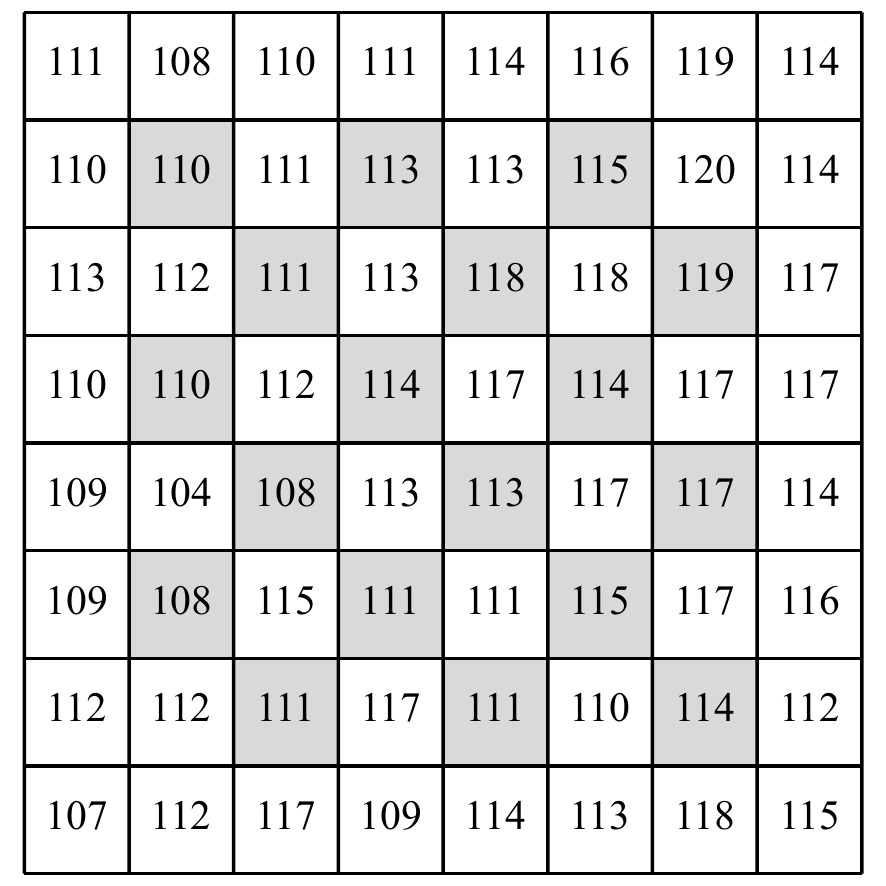}
}
\subfigure[]{
\includegraphics[scale=0.7]{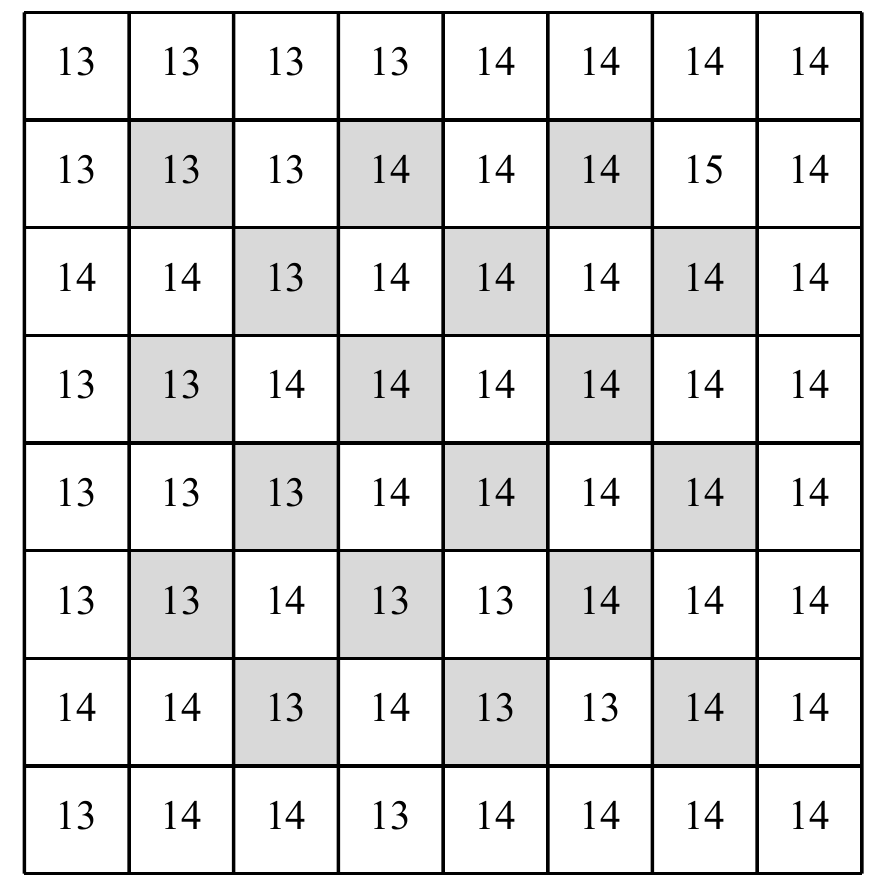}
}
\quad
\caption{\textrm{Randomly select a pixel block with size $8\times8$ from Lena and its corresponding pixel value matrix: (a) Selected pixel block; (b) MSB matrix with $n=3$.}}
\vspace{-0.5cm}
\end{figure}

\begin{figure}[htbp]
\centering
	\includegraphics[scale=0.7]{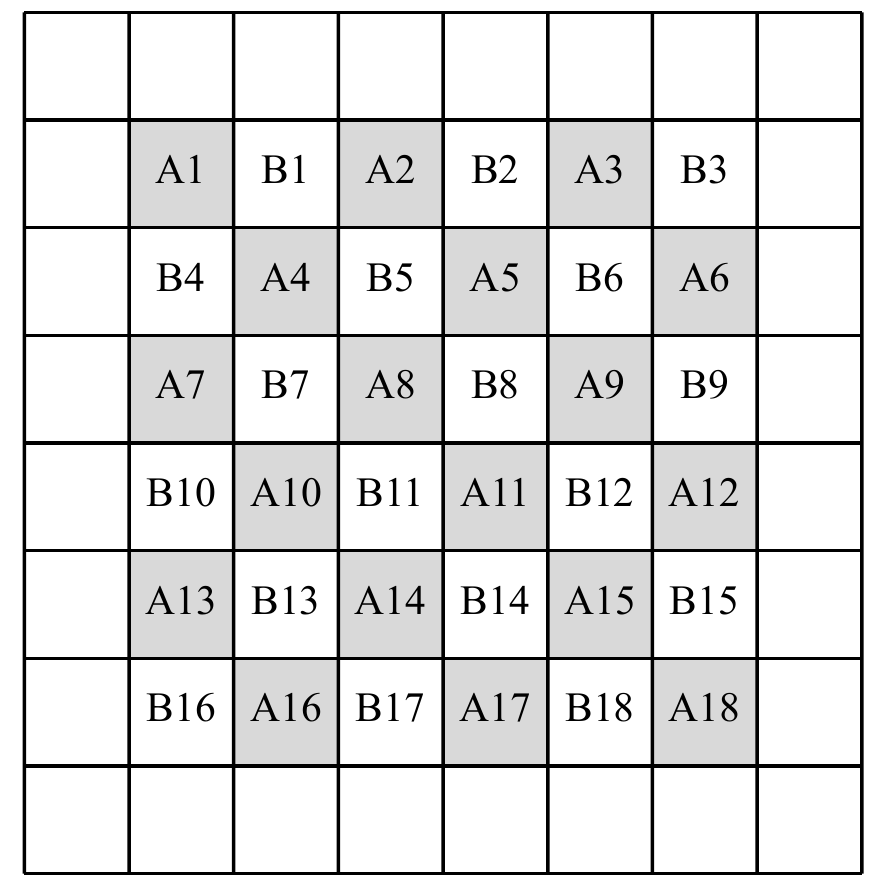}
    \caption{\centering \textrm{The pixel number of grey pixels.}}
    \vspace{-0.5cm}
\end{figure}

\begin{figure}[htbp]
\centering
\subfigure[]{
\includegraphics[width=5cm]{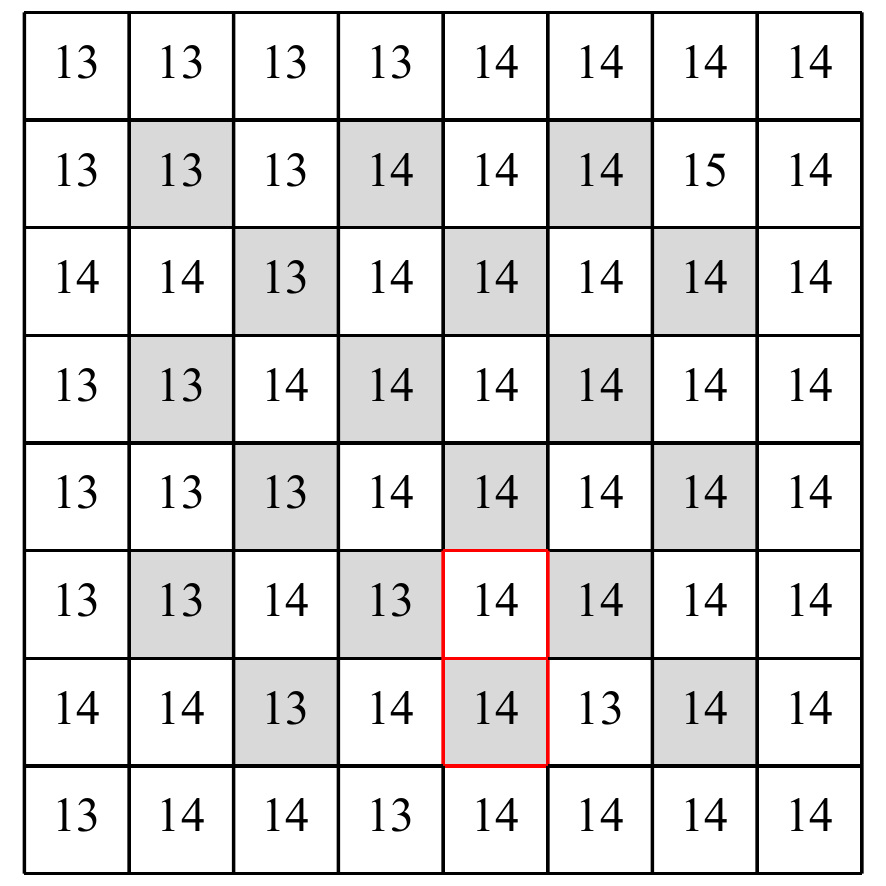}
}
\subfigure[]{
\includegraphics[width=5cm]{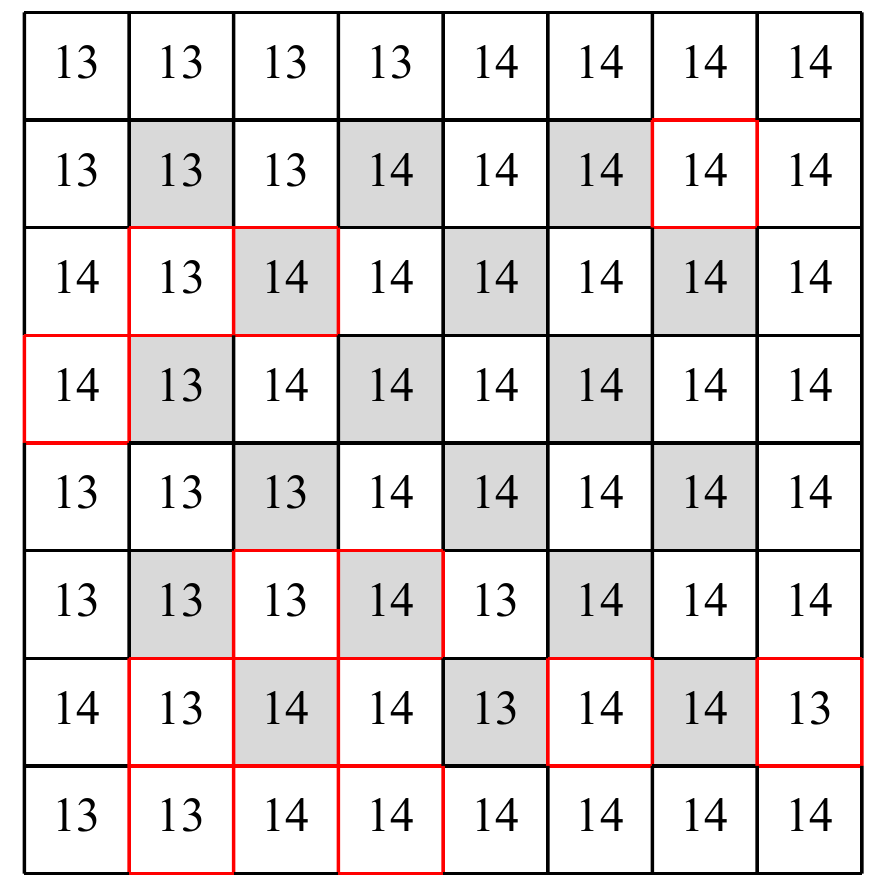}
}
\subfigure[]{
\includegraphics[width=5cm]{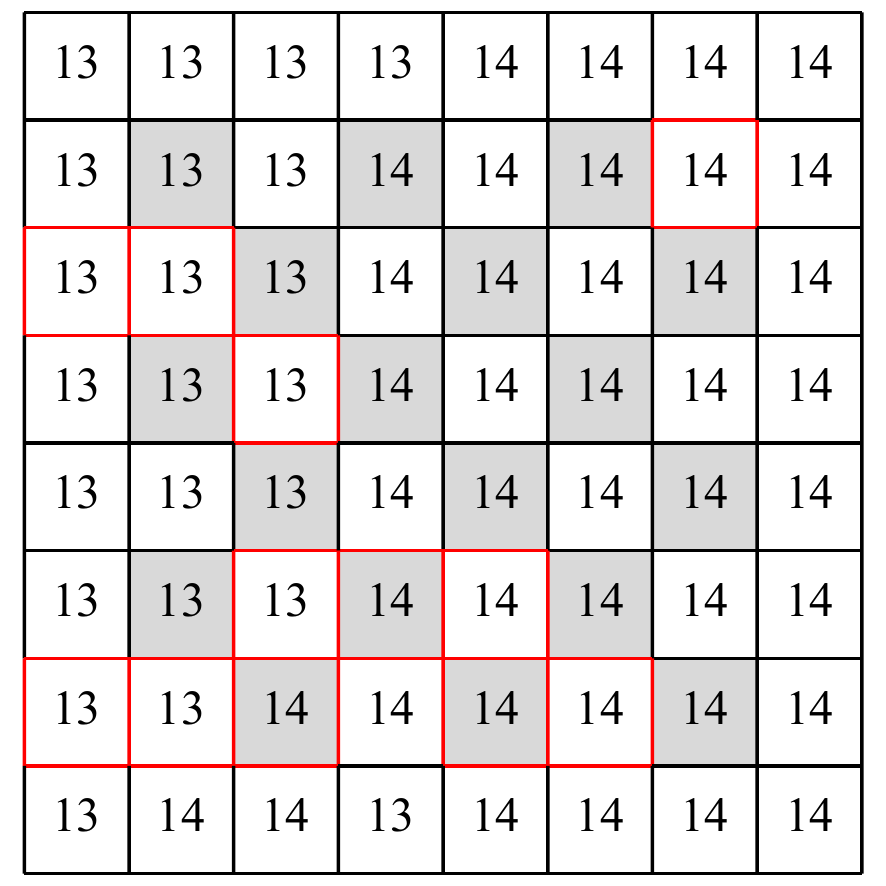}
}
\quad
\caption{\centering \textrm{MSB matrix of compressed pixel blocks: (a) QF = 100; (b) QF = 90; (c) QF = 85.}}
\vspace{-0.5cm}
\end{figure}

Because the processing of white pixels is similar to that of grey pixels, only grey pixels are taken as an example here to calculate the local complexity and observe the pixel value ordering changes before and after compression. To better illustrate the local complexity of each grey pixel, we number it as shown in \textcolor{blue}{Fig. 8}.

Original pixel blocks are compressed with different quality factors, e.g., QF = 100, 90, 85. Then the compressed pixel blocks are divided into two parts, namely LSB and MSB matrix under $n=3$. MSB matrixes under different quality factors are shown in \textcolor{blue}{Fig. 9}. The pixels in red boxes in \textcolor{blue}{Fig. 9} indicate the pixels that have been changed after JPEG compression. We calculate the local complexity of the uncompressed and compressed MSB grey pixels, which is shown in \textcolor{blue}{Table 3}.

\begin{table}
\caption{\textrm{Changes in local complexity of grey pixels because of JPEG compression:(a) grey pixels A1-A9; (b) grey pixels A10-A18.}} \label{tbl1}
\begin{tabular*}{\tblwidth}
{@{}p{1.22cm}p{1.29cm}p{1.29cm}p{1.29cm}p{1.29cm}p{1.29cm}p{1.29cm}p{1.29cm}p{1.29cm}p{1.29cm}@{}}
\toprule
\textrm{QF}& \textrm{A1} & \textrm{A2} & \textrm{A3} &\textrm{A4}& \textrm{A5} & \textrm{A6} & \textrm{A7} & \textrm{A8} & \textrm{A9} \\
\midrule
\textrm{Original} & \textrm{0.1875} & \textrm{0.2500} & \textrm{0.1875}& \textrm{0.1875}&	\textrm{0}	& \textrm{0.1875} &	\textrm{0.2500}	& \textrm{0}& \textrm{0}	\\

\textrm{100}& \textrm{0.1875}&	\textrm{0.2500}& 	\textrm{0.1875}& \textrm{0.1875} &	\textrm{0} &	\textrm{0.1875}& \textrm{0.2500} &	\textrm{0} &\textrm{0} \\

\textrm{90} &\textrm{0.1875}	&\textrm{0}&	\textrm{0.2500}&	\textrm{0}&	\textrm{0}&\textrm{0.2500}&\textrm{0}&	\textrm{0}	&\textrm{0.2500} \\
\textrm{85} & \textrm{0} &\textrm{0.1875} &\textrm{0}&\textrm{0.1875}&	\textrm{0}	&\textrm{0}	&\textrm{0}&	\textrm{0.1875}&\textrm{0}	\\
\bottomrule
\end{tabular*}\\
\textrm{(a)}\\
\begin{tabular*}
{\tblwidth}
{@{}p{1.22cm}p{1.29cm}p{1.29cm}p{1.29cm}p{1.29cm}p{1.29cm}p{1.29cm}p{1.29cm}p{1.29cm}p{1.29cm}@{}}
\toprule
\textrm{QF}&  \textrm{A10}& \textrm{A11} &\textrm{A12} & \textrm{A13} & \textrm{A14} & \textrm{A15} & \textrm{A16} & \textrm{A17} &\textrm{A18} \\
\midrule
\textrm{Original} & \textrm{0.1875} &	\textrm{0.1875} & \textrm{0}	&\textrm{0.2500}	& \textrm{0.1875}&	\textrm{0.2500}&\textrm{0}	& \textrm{0.2500}&	\textrm{0.1875} \\

\textrm{100} &\textrm{0.1875} & \textrm{0} &\textrm{0}	 &\textrm{0.2500}& \textrm{0}	& \textrm{0.1875} &\textrm{0} &\textrm{0.1875} & \textrm{0.1875} \\

\textrm{90} &\textrm{0.1875	}&\textrm{0}&	\textrm{0}	&\textrm{0.2500	}&\textrm{0.1875}	&\textrm{0.2500}&	\textrm{0.1875}&	\textrm{0.1875}&	\textrm{0.1875}\\
\textrm{85} 	&\textrm{0.1875}&\textrm{0}&\textrm{0}&	\textrm{0}&\textrm{0.1875}&\textrm{0}&\textrm{0.2500}	&\textrm{0}	&\textrm{0}\\
\bottomrule
\end{tabular*}\\
\textrm{(b)}
\end{table}

We arrange the pixels according to the ascending order of their local complexity. From \textcolor{blue}{Table 3}, the order of uncompressed pixel block in \textcolor{blue}{Fig. 7(b)} would be O = (A5, A8, A9, A12, A16, A1, A3, A4, A10, A14, A18, A2, A13, A15, A17). When pixel block is compressed with QF = 100 (\textcolor{blue}{Fig. 9(a)}), it is found that the value of (A17, B14) changes from (13, 13) to (14, 14). As the value of white pixel B14 changes, the local complexity of the surrounding pixels (A11, A14, A15, A17) changes from (0.1875, 0.1875, 0.2500, 0.2500) to (0, 0, 0.1875,0.1875), which results in a change in pixel ordering. As a result, the order of compressed pixel under QF = 100 is changed to O$_{100}$ = (A5, A8, A9, A11, A12, A14, A16, A1, A3, A4, A6, A10, A15, A17, A18, A2, A7, A18, A2, A7, A13). There are only A5, A8, A9 still in the same place. The order of compressed pixel under QF = 90 is changed to O$_{90}$ = (A1, A3, A5, A6, A8, A9, A12, A11, A15, A17, A18, A4, A7, A10, A14, A16). None of the pixels is in their original position. The order of compressed pixel under QF = 85 is changed to O$_{85}$ = (A1, A3, A5, A6, A7, A9, A11, A12, A13, A15, A17, A18, A2, A4, A8, A10, A14, A16). As with a QF of 85, all the pixels are out of place. Since JPEG compression causes changes in pixel value, and the local complexity  of pixels depends on the pixel value, JPEG compression will  lead to errors in calculating local complexity, resulting in confusion in pixel value ordering. As a result, the secret data extracted by the receiver is grossly inconsistent with the original secret data.

\section{Proposed suggestions on improving the robustness}

On the premise of ensuring reversibility, to improve the robustness of the scheme \textcolor{blue}{\cite{kumar2020robust}}, we give some suggestions. The first two suggestions are made for two robust testings respectively. The third suggestion is made for the prediction error expansion-based embedding method in the two-layer embedding strategy.

\textbf{Strategy 1: Adopting spread spectrum technology \textcolor{blue}{\cite{hartung1998watermarking}} to embed auxiliary information and using majority voting system to extract.} Spread spectrum communication schemes use frequency spreading to send a narrow-band signal over a wide-band channel,  substantially improving the robustness of the signal  \textcolor{blue}{\cite{hartung1998watermarking}}. The majority voting system refers to voting on events and selecting the majority result as the correct result. According to the first robustness testing in Section 4, the auxiliary information is easily tampered with. In order to ensure the accuracy of auxiliary information extraction, we suggest that the data-hider uses spread spectrum technology to embed each bit of the auxiliary information repeatedly (at least 3 times) to obtain the extended auxiliary information. The receiver uses the majority voting system to select the majority of the information as the extracted information. In this way, even if the marked image is tempered with, as long as half of the extended auxiliary information can be extracted correctly, the auxiliary information bit can be guaranteed to be correct, which greatly improves the robustness of the scheme \textcolor{blue}{\cite{kumar2020robust}}.

\textbf{Strategy 2: Embedding secret data in raster scan order.} From the second robust testing shown in Section 4, it can be seen that if the pixel values are sorted according to the local complexity, the sorting will be severely disrupted when the marked image is attacked, which leads to serious inconsistencies between the secret data extracted by the receiver and the original secret data. Therefore, to avoid this situation, we recommend embedding secret data directly in raster scan order (left to right and top to bottom of the image). The receiver also extracts the secret data in raster scan order, which avoids the confusion of the order of pixel values sorted by local complexity caused by JPEG compression, thereby improving the robustness.

\textbf{Strategy 3: Increasing the shift quantity of prediction error histogram.} From Step 5 in Section 2, it can be seen that the shift quantity of prediction error histogram adopted by the scheme \textcolor{blue}{\cite{kumar2020robust}} is +1 or -1. In this embedding mode, as long as the pixel value is slightly changed, the receiver is likely to extract the wrong prediction error, thus leading to the extraction of wrong secret data. Therefore, we recommend using a larger shift quantity referring to the scheme \textcolor{blue}{\cite{zeng2010lossless}} to create a robust region to accommodate the error caused by the pixel value change after JPEG compression. Due to the existence of the robust region, when the marked image is attacked to a certain extent, if the pixel value changes in a small range, as long as its predictor error has not been tempered with into the wrong region, the secret data can still be extracted correctly. In this way, the robustness of the scheme \textcolor{blue}{\cite{kumar2020robust}} can be further improved.

\section{Conclusion}

In the paper ``Robust reversible data hiding scheme based on two-layer embedding strategy'' published in INS recently, Kumar et al. proposed an RRDH scheme based on two-layer embedding. However, JPEG compression leads to a change in MSB planes, which leads to auxiliary information being damaged, which in turn leads to a change in the predictor. As a result, the secret data extracted by the receiver will be significantly different from the original secret data. The damaged pixel values caused by JPEG compression change the pixel value ordering, so the receiver extracts the secret data in the wrong order. From the above analysis, the secret data cannot be accurately extracted from compressed images, so Kumar et al.'s scheme is not robust. Finally, possible suggestions have been provided to improve the robustness of their scheme.\par

\section*{Declaration of competing interest}
This manuscript is the author's original work and has not been published nor has it been submitted simultaneously elsewhere.

\section*{Acknowledgements}
The authors thank the anonymous referees for their valuable comments and suggestions. This research work is partly supported by National Natural Science Foundation of China (61872003, U20B2068, 61860206004).

\bibliographystyle{unsrt}
\bibliography{reference}

\end{document}